\documentclass[prd,superscriptaddress,twocolumn,showpacs,amsmath,amssymb]{revtex4}
\usepackage{amsmath}
\usepackage{graphicx}
\usepackage{color}
\usepackage{dcolumn}
\usepackage{bm}
\usepackage{slashed}

\newcommand{\bea}{\begin{eqnarray*}}
\newcommand{\eea}{\end{eqnarray*}}
\newcommand{\bean}{\begin{eqnarray}}
\newcommand{\eean}{\end{eqnarray}}
\newcommand{\eqs}[1]{Eqs.(\ref{#1})}
\newcommand{\eq}[1]{Eq.(\ref{#1})}
\newcommand{\meq}[1]{(\ref{#1})}
\newcommand{\grad}{\nabla}
\newcommand{\eqn}{&=&}
\newcommand{\non}{\nonumber \\}
\newcommand{\sgt}{\sqrt{h}}

\newcommand{\hsp}{\hspace{0.1mm}}
\newcommand{\rt}{R^{(3)}}

\newcommand{\pp}{\partial}

\begin{document}
\title{Thermodynamical stability for perfect fluid}

\author{Xiongjun Fang}
\email[Xiongjun Fang: ]{fangxj@hunnu.edu.cn} \affiliation{Department of Physics, Key Laboratory of Low Dimensional Quantum Structures and
Quantum Control of Ministry of Education, and Synergetic Innovation Center for Quantum Effects and Applications, Hunan Normal
University, Changsha, Hunan 410081, P. R. China}

\author{Xiaokai He}
\email[Xiaokai He: ]{hexiaokai77@163.com} \affiliation{Department of Physics, Key Laboratory of Low Dimensional Quantum Structures and Quantum Control of Ministry of Education, and Synergetic Innovation Center for Quantum Effects and Applications, Hunan Normal University, Changsha,
Hunan 410081, P. R. China}

\affiliation{School of Mathematics and Computational Science, Hunan First Normal University, Changsha 410205, China}

\author{Jiliang Jing}
\email[corresponding author: Jiliang Jing, ]{ jljing@hunn.edu.cn} \affiliation{Department of Physics, Key Laboratory of Low Dimensional Quantum Structures and
Quantum Control of Ministry of Education, and Synergetic Innovation Center for Quantum Effects and Applications, Hunan Normal
University, Changsha, Hunan 410081, P. R. China}

\begin{abstract}
According to maximum entropy principle, it has been proved that the gravitational field equations could be derived by the extrema of total entropy for perfect fluid, which implies that thermodynamic relations contain information of gravity. In this manuscript, we obtain a criterion for thermodynamical stability of an adiabatic, self-gravitating perfect fluid system by the second variation of total entropy. We show, for Einstein's gravity with spherical symmetry spacetime, that the criterion is consistent with that for dynamical stability derived by Chandrasekhar and Wald. We also find that the criterion could be applied to cases without spherical symmetry, or under general perturbations. The result further establishes the connection between thermodynamics and gravity.
\end{abstract}


\maketitle

\section{Introduction}
The connection between thermodynamics and gravity has attracted great attention in recent decades. In 1994, Jacobson showed that Einstein equations could be derived from fundamental thermodynamic relation which is hold for local Rindler causal horizons \cite{Jacobson1}. Verlinder put forward an interesting but incomplete viewpoint that gravity could be explained as entropy force \cite{Verlinder}. Meanwhile, a series of papers proved that the maximum entropy principle for perfect fluid in different theories. This principle shows that if the constraint equation and some thermodynamic conditions are satisfied, the gravitational field equations could be derived by the extrema of total entropy \cite{wald1981,gao,fang1,fang2,fang3,Cao1,Cao2}. Recently, Jacobson proposed a structure ``causal diamond'' and considered that the entanglement equilibrium would imply the Einstein equations \cite{Jacobson2}. All of these studies suggest a general and solid connection between thermodynamics and gravity. Moreover, the emergence has also been widely discussed in the past years \cite{Emerge1,Emerge2}. It was considered that the gravity may not be the fundamental assumption of our nature. However, most of these discussions about the connection between gravity and thermodynamics are focused on establishing the relation between the first order variation of thermodynamical properties and the gravitational equations. This manuscript will reveals their connection in higher order variation by investigating the stability of perfect fluid in static background spacetime.

In general there are two methods of testing the stability of a static configuration of fluid \cite{Cocke}. One is dynamical method, and the other is thermodynamical method. In dynamical method, it requires that the physical quantities deviated only slightly from equilibrium state. Assuming that the motions are adiabatic and reversible, then the variation separation approach makes the perturbational fields $\psi^a$ takes the form $\psi^a(r)e^{-i\omega t}$. Then the equation for $\psi^a$ could be transformed to a Sturm-Liouville eigenvalue problem, with $\omega$ the eigenvalue. The stability is tested by ascertaining whether or not all $\omega^2$'s are positive. The stability problem in general relativity was first discussed by Chandrasekhar \cite{Chandrasekhar1,Chandrasekhar2}. After that, Friedman investigated the dynamic stability of relativistic stars with respect to perturbations that arise in the Lagrangian displacement framework \cite{Friedman1,Friedman2,Friedman3}. Recently, Wald and Seifert presented a general dynamical method for the analysis of the stability of static, spherically symmetric solutions to spherically symmetric perturbations \cite{waldgeneral}. This method could be used in an arbitrary diffeomorphism covariant Lagrangian field theory in which the field equations are at most second order derivatives of the metric \cite{Seifert}. In thermodynamical method, to isolated system, the negative of the second variation of total entropy corresponding to thermodynamical stability, which was discussed by Cocke \cite{Cocke} and Sorkin \cite{wald1981}. Recently, Roupas proved that the maximum of total entropy for perfect fluid gives the same criterion as Yabushita's result with some additional conditions \cite{Roupas,Yabu}.

However, using dynamical method to solve stability problem in static spacetime always be limited within the spherical symmetry and radial perturbation. For the cases without spherical symmetry, or with more general perturbations, dynamical method is very hard to deal with. We believe that thermodynamical method is promising to solve these difficulty. In this manuscript, by extending Roupas' result, we obtain a general formula of the second variation of the total entropy for perfect fluid, as the criterion for thermodynamical stability. It should be noted that this criterion could be applied in general static background without spherical symmetry.

It is worth noting that recently Wald et al. presented a very comprehensive discussion on the equivalence of dynamical stability and thermodynamical stability \cite{wald2013}. The results in \cite{wald2013} seems to be similar to a part of our manuscript, however, our assumptions and arguments are different from \cite{wald2013} . For instance, with the definition of ADM mass, a crucial assumption in \cite{wald2013} is that the spacetime should be asymptotically flat, while our derivations apply to any region imposing no global conditions on spacetime.

The rest of this paper is organized as follows. In the next section, we give a formula as the thermodynamical stability criterion for perfect fluid in static background spacetime firstly. Then this criterion was applied specifically to the case of radial perturbations of static, spherical symmetric perfect fluid in Einstein gravity. It was found that our thermodynamical criterion is consistent with the dynamical criterion found by Wald. In section III, the explicit expression of the thermodynamical stability criterion of perfect fluid under non-radial perturbation in general static background was obtained. At last, we summarized this paper with some comments and discussions.

Throughout our discussion, units will be used in which $c=G=1$. The letters (a, b, c) denote the abstract index. We also ignore the factor $\kappa=8\pi$ in Einstein equations, hence $G_{\mu\nu}=T_{\mu\nu}$.

\section{Criterion for Thermodynamical Stability}

First, we briefly review the proof of the maximum entropy principle for perfect fluid in general static spacetime \cite{fang1}. We assume that all the quantities are measured by static background observers which are orthogonal to the hypersurface $\Sigma$, and we consider that the fluid over any selected region $C$ on $\Sigma$ satisfied ordinary thermodynamic relations and Tolman's law, which gives $T\chi=const.$, where $T$ and $\chi$ are the temperature of the fluid and the redshift factor, respectively. Without loss of generality, we take $T\chi=1$. In \cite{fang1} we have shown that if the constraint equation and some boundary conditions are satisfied, and the total particle number is fixed, the variation of the total entropy is
\bean \label{deltaS}
\delta S = \int_C\frac{\sgt}{T}\delta\rho+\frac{p+\rho}{T}\delta\sgt \,,
\eean
where $\rho$ and $p$ represent the energy density and the pressure of the fluid, respectively, and $h$ is the determinant of the induced metric $h_{ab}$ on $\Sigma$. Then we proved that the extrema of the total entropy $\delta S=0$ gives the components of gravitational field equations \cite{fang1}. The maximum entropy principle implies that the gravitational field equations may be replaced by constraint equation and thermodynamic relations.

A isolated star in thermodynamical equilibrium is said to be thermodynamical stable if $\delta^2S<0$. When discuss about the second variation of the total entropy, we consider that the Tolman's law is also valid since the state is deviated slightly from equilibrium state. Applying the local first law of thermodynamics
\bean
T\delta s=\delta \rho-\mu \delta n \,,
\eean
and the integrated form of the Gibbs-Duhem relation
\bean
p+\rho=Ts+\mu n \,,
\eean
with the fact that $\mu/T=const.$ \cite{fang1}, we obtain
\bean
\delta p=s\delta T+n\frac{\mu}{T}\delta T=\frac{p+\rho}{T}\delta T \,,
\eean
where $\mu$ and $n$ denote chemical potential and particle number density, respectively. So the second variation of total entropy could be written as
\bean \label{delta2S}
\delta^2S \eqn \int_C\frac{1}{T}\left[2\delta\rho\delta\sgt+\sgt\delta^2\rho \right. \non
&& \left. +(p+\rho)\delta^2\sgt-\frac{\delta p\delta\rho}{p+\rho}\sgt\right]\,.
\eean
And $\delta^2S<0$ means the system is thermodynamical stable. Hence, using \eq{delta2S} one could directly obtain the specific form of stability criterion. A natural question is that whether the thermodynamical stability is equivalent to dynamical stability or not.

As a concrete example, we investigate whether this equivalence is valid for spherically symmetric perturbations of static, spherical symmetric perfect fluid in Einstein gravity. For such spacetime, the metric takes the form \cite{MTW}
\bean \label{metric}
ds^2=-e^{2\Phi(t,r)}dt^2+e^{2\Lambda(t,r)}dr^2+r^2d\Omega^2 \,.
\eean
According to Chandrasekhar's procedure, under the perturbation, the four-velocity becomes \cite{Chandrasekhar2}
\bean
u_0=-e^{\Phi}\,, && u^0=e^{-\Phi} \,, \non
u_1=e^{2\Lambda-\Phi}\vec v\,, && u^1=e^{-\Phi}\vec v \,,
\eean
where $\vec v=dr/dt=\pp\xi/\pp t$. Here $\xi$ is the radial ``Lagrangian displacement", which describes the radial displacement of each fluid element from its ``equilibrium position". Then the $t-r$ component of Einstein equations, i.e. $T_0\hsp^1=G_0\hsp^1$, gives
\bean
-(p+\rho)e^{2\Lambda}\frac{\pp \xi}{\pp x^0}=\frac{2}{r}\frac{\pp\Lambda}{\pp x^0} \,.
\eean
Direct integration yields
\bean
\delta\Lambda=-\frac{r}{2}e^{2\Lambda}(p+\rho)\xi \,.
\eean
With the first variation of constraint equation, $\delta G_{00}=\delta T_{00}$, we have
\bean
\delta\rho=-\frac{1}{r^2}\frac{\pp}{\pp r}[r^2(p+\rho)\xi] \,.
\eean
So the second variation of $\rho$ could be written as
\bean
\delta^2\rho = -\frac{1}{r^2}\frac{\pp}{\pp r}[r^2(\delta p+\delta \rho)\xi+r^2(p+\rho)\delta\xi] \,.
\eean
Meanwhile, the variation of the induced metric could be written as
\bean
\delta\sgt = e^{\Lambda}r^2\sin\theta\cdot \delta\Lambda = -\frac{1}{2}e^{3\Lambda}r^3(p+\rho)\sin\theta\cdot\xi \,,
\eean
and
\bean
\delta^2\sgt \eqn\frac{3}{4}r^4e^{5\Lambda}(p+\rho)^2 \sin\theta \cdot \xi^2 \non
&& -\frac{1}{2}e^{3\Lambda}r^3(\delta p+\delta\rho)\sin\theta \cdot\xi \non
&& -\frac{1}{2}e^{3\Lambda}r^3(p+\rho)\sin\theta \cdot\delta\xi \,.
\eean

From \eq{metric}, $1/T=\chi=\sqrt{-g_{00}}=e^{\Phi}$, using integration by parts and drop the boundary terms, the first three terms in the right hand side of \eq{delta2S} could be calculated one by one. Note that in this case $\int_C$ becomes $4\pi\int_rdr$. Explicitly, the first term becomes
\begin{widetext}
\bean \label{first}
4\pi\int_r dr\frac{2}{T}\delta\rho\delta\sgt
\eqn 4\pi\int_r dr \frac{e^{\Phi+3\Lambda}}{2r}\frac{\pp}{\pp r}[r^2(p+\rho)\xi]^2 \non
\eqn 4\pi\int_r dr\left[-\frac{3}{4}e^{\Phi+5\Lambda}r^4(p+\rho)^3\xi^2
+\frac{1}{2}e^{\Phi+3\Lambda}\left(\frac{2\Phi'}{r}+\frac{1}{r^2}\right)r^4(p+\rho)^2\xi^2 \right] \,,
\eean
the second term becomes
\bean \label{second}
4\pi\int_r dr\frac{1}{T}\sgt\delta^2\rho
\eqn 4\pi\int_r dr e^{\Phi+\Lambda}r^2 \cdot \frac{1}{r^2}\frac{\pp}{\pp r}[r^2(\delta p+\delta \rho)\xi+r^2(p+\rho)\delta\xi] \non
\eqn 4\pi \int_r dr e^{\Phi+3\Lambda}\left[\frac{1}{2}r^3(p+\rho)(\delta p+\delta \rho)\xi+\frac{1}{2}r^3(p+\rho)^2\delta\xi\right] \,,
\eean
and the third term becomes
\bean \label{third}
&& 4\pi\int_r dr\frac{1}{T}(p+\rho)\delta^2\sgt=4\pi\int_r dr\left[\frac{3}{4}r^4e^{\Phi+5\Lambda}(p+\rho)^3\xi^2-\frac{1}{2}e^{\Phi+3\Lambda}r^3(\delta p+\delta\rho)(p+\rho)\xi
-\frac{1}{2}e^{\Phi+3\Lambda}r^3(p+\rho)^2\delta\xi \right] \,.\non
\eean
Substituting these results into \eq{delta2S} we obtain
\bean \label{delta2S2}
\delta^2S = 4\pi\int_r dr\left\{\frac{1}{2}e^{\Phi+3\Lambda}\left(\frac{2\Phi'}{r}+\frac{1}{r^2}\right)r^4(p+\rho)^2\xi^2
-e^{\Phi+\Lambda}r^2\frac{\delta p\delta \rho}{p+\rho}\right\}
\eean
\end{widetext}
Note that we have chosen the energy density $\rho$ and the particle number density $n$ as independent variables when obtained the first variation of total entropy \cite{fang1}. But Chandrasekhar has chosen $\rho$ and $p$ as independent variables \cite{Chandrasekhar1,Chandrasekhar2}. It is not obviously to compare our result to Chandrasekhar's. However, using dynamical method, Seifert and Wald give a stability criterion of the star with a ``barotropic" equation of the state of the form $\rho=\rho(n)$, in this situation there is only one thermodynamical variable. We will show that our stability criterion would reduce to Wald's result exactly. For this purpose, we start with the Lagrangian for the perfect fluid used in \cite{waldgeneral}
\bean \label{Lmatt}
\mathcal{L}_{mat}=-\varrho(n) \,.
\eean
Wald showed that there exist an identification:
\bean
\rho \rightarrow \varrho \,, ~\ p \rightarrow \frac{\pp\varrho}{\pp n}n-\varrho \,.
\eean
From which it is easy to obtain
\bean
\delta\rho = \frac{\pp\varrho}{\pp n}\delta n \,, ~\ \delta p = \frac{\pp^2\varrho}{\pp n^2}n\delta n \,.
\eean
\begin{widetext}
So \eq{delta2S2} becomes
\bean
\delta^2 S = 4\pi\int_r dr\left[r^2e^{\Phi+\Lambda}\left(2\frac{\pp\Phi}{\pp r}+\frac{1}{r}\right)\left(\frac{\pp\Lambda}{\pp r}
+\frac{\pp\Phi}{\pp r}\right)(p+\rho)\xi^2-e^{\Phi+\Lambda}r^2\frac{\pp^2\varrho}{\pp n^2}(\delta n)^2\right] \,.
\eean
Note that the variation of particle number density $n$ could be written as \cite{waldgeneral}
\bean
\delta n = n\left[\frac{\pp\xi}{\pp r}+\left(\frac{1}{\nu}\frac{\pp\nu}{\pp r}+\frac{\pp\Lambda}{\pp r}+\frac{2}{r}\right)\xi
-\left(\frac{\pp\Phi}{\pp r}+\frac{\pp\Lambda}{\pp r}\right)\xi\right]=\frac{e^{\Phi}}{r^2}\frac{\pp}{\pp r}(r^2e^{-\Phi}n \xi) \,.
\eean
We obtain
\bean \label{delta2SGR}
\delta^2S = 4\pi\int_r dr \left[r^2e^{\Phi+\Lambda}\left(2\frac{\pp\Phi}{\pp r}+\frac{1}{r}\right)
\left(\frac{\pp\Lambda}{\pp r}+\frac{\pp\Phi}{\pp r}\right)(p+\rho)\xi^2-\frac{e^{3\Phi+\Lambda}}{r^2}\frac{\pp^2\varrho}{\pp n^2}
\left(\frac{\pp}{\pp r}(r^2e^{-\Phi}n\xi)\right)^2\right] \,.
\eean
\end{widetext}
Note that the thermodynamical stability requires only the second variation of total entropy be negative. The above expression \eq{delta2SGR} agrees with the result in Seifert and Wald. So far, we have shown that our criterion for thermodynamical stability is consistent with the dynamical stability criterion for spherical perturbations of a static, spherical symmetry star with a barotropic equation of state. In next section, we will give the stability criterion in more general cases.

\section{Thermodynamical stability criterion in general cases without spherical symmetry}
The criterion for thermodynamical stability applied not only for the above particular case, but also for more general case. In static spacetime, for the cases without spacial symmetry, or with non-radial perturbations, the method of dynamical stability is hard to deal with them. However, thermodynamical stability seems more easily to handle these cases. Assuming that the metric of perfect fluid star in background spacetime could be written as
\bean
ds^2=g_{00}dt^2+g_{ij}dx^idx^j \,,
\eean
and the perturbation fields are $\delta g_{\mu\nu}$, we show how to get the explicitly form of the second variation of total entropy $\delta^2S$ in general static spacetime. In this section, $h_{ab}$ denote the induced metric of the $t=const.$ slice $\Sigma$. And we use $A_a$, $D_a$ and $\Box$ denote the four-acceleration of the observer, the 3-dim covariant derivative and 3-dimensional Box operator $D_aD^a$ in $\Sigma$, respectively.

The extrinsic curvature of $\Sigma$ is
\bean
K_{ab}=h_a\hsp^c\grad_c u_b \,.
\eean
The relation between the ordinary curvature $R$ and the 3-dimensional curvature $\rt$ on $\Sigma$ is given by
\bean \label{rtR}
\rt=R+2R_{ab}u^au^b-\frac{1}{2}(K_a\hsp^a)^2+\frac{1}{2}K_{ab}K^{ab} \,,
\eean
which yields
\bean \label{rhodeviated}
\rho=\frac{1}{2}\rt+\frac{1}{2}(K_a\hsp^a)^2-\frac{1}{2}K_{ab}K^{ab} \,.
\eean
Note that $K_{ab}|\hsp_0=0$ in static background spacetime. It is obtained that
\bean
\delta\rho=\frac{1}{2}\delta \rt \,,
\eean
and
\bean \label{delta2rho}
\delta^2\rho \eqn \frac{1}{2}\delta^2\rt+h^{ac}h^{bd}(\delta K_{ac}\delta K_{bd}-\delta K_{ab}\delta K_{cd}) \,.
\eean

The perturbation of the induced metric is
\bean
\delta\sgt=\frac{1}{2}\sgt h^{ab}\delta h_{ab} \,.
\eean
Then we calculate the variation of extension curvature, $\delta K_{ab}$,
\bean
\delta K_{ab} \eqn \delta (h_a\hsp^c\grad_cu_b) \non
\eqn \grad_cu_b\delta h_a\hsp^c+h_a\hsp^c\grad_c\delta u_b-h_a\hsp^c\delta\Gamma^{\mu}\hsp_{cb} u_{\mu} \,.
\eean
Using
\bean
\delta u_a=-\frac{1}{2}u_au^bu^c\delta g_{bc} \,,
\eean
and the fact $\grad_au_b=-A_bu_a$ in static background, we have
\bean \label{KK1}
h^{ac}h^{bd}\delta K_{ac}\delta K_{bd}=h^{ab} h^{cd}u_{e}u_{f}\delta\Gamma^{e}\hsp_{ab}\cdot \delta\Gamma^{f}\hsp_{cd} \,,
\eean
and
\bean \label{KK2}
-h^{ac}h^{bd}\delta K_{ab}\delta K_{cd}=-h^{ac}h^{bd}u_{e}u_{f}\delta\Gamma^{e}\hsp_{ab}\cdot \delta\Gamma^{f}\hsp_{cd} \,.
\eean
Substituting \eqs{KK1} and \meq{KK2} into \eq{delta2rho}, the expression of $\delta^2\rho$ could be obtained.

Now we calculate each terms in \eq{delta2S} one by one. Noting that the induced metric and its derivatives are fixed on the boundary of the selected region $C$, we could use integration by parts and drop the boundary terms. The first term of \eq{delta2S} could be written as
\begin{widetext}
\bean \label{genefirst}
\mathcal{G}_1 \eqn \int_C \frac{2}{T}\delta\rho\delta\sgt \non
\eqn \int_C\frac{\chi}{2}\sgt \left( h^{cd}\delta h_{cd}\cdot D^aD^b\delta h_{ab}
- h^{ab}h^{cd}\delta h_{cd}\cdot D^eD_e\delta h_{ab}-h^{cd}\rt\hsp^{ab}\delta h_{ab}\delta h_{cd} \right) \,.
\eean
The second term of \eq{delta2S} becomes
\bean
\mathcal{G}_2 \eqn \int_C \frac{\sgt}{T} \delta^2\rho \non
\eqn \int_C \frac{\chi\sgt}{2}\delta\left(h^{ab}\delta\rt_{ab}+\rt_{ab}\delta h^{ab}\right)
+\chi\sgt (h^{ab}h^{cd}-h^{ac}h^{bd}) u_{\mu}u_{\nu}\delta\Gamma^{\mu}\hsp_{ab}\cdot\delta\Gamma^{\nu}\hsp_{cd} \non
\eqn \int_C \frac{\chi\sgt}{2}\delta \left[h^{ac}h^{bd}D_c D_d\delta h_{ab}-h^{ab}D^cD_c\delta h_{ab}+\rt_{ab}\delta h^{ab} \right] \non
&& +\chi\sgt (h^{ab}h^{cd}-h^{ac}h^{bd}) u_{e}u_{f}\delta\Gamma^{e}\hsp_{ab}\cdot\delta\Gamma^{f}\hsp_{cd} \non
\eqn \int_C \frac{\chi\sgt}{2} \left[h^{bd}\delta h^{ac}\cdot D_c D_d\delta h_{ab}+h^{ac}\delta h^{bd}\cdot D_c D_d\delta h_{ab} +h^{ac}h^{bd}\delta(D_c D_d\delta h_{ab}) \right. \non
&& \left. -\delta h^{ab}\cdot D^cD_c\delta h_{ab}-h^{ab}\delta(D^cD_c\delta h_{ab})+\delta \rt_{ab}\delta h^{ab}+\rt_{ab}\delta^2h^{ab} \right] \non
&& +\chi\sgt (h^{ab}h^{cd}-h^{ac}h^{bd}) u_{e}u_{f}\delta\Gamma^{e}\hsp_{ab}\cdot\delta\Gamma^{f}\hsp_{cd} \,,
\eean
where $h^{ac}h^{bd}\delta(D_c D_d\delta h_{ab})$ and $-h^{ab}\delta(D^cD_c\delta h_{ab})$ could be calculated as
\bean
&& h^{ac}h^{bd}\delta(D_c D_d\delta h_{ab}) \non
\eqn h^{ac}h^{bd}D_c D_d\delta^2 h_{ab}-\frac{1}{2}\Box\delta h^{ab}\cdot \delta h_{ab}+D_aD^b\delta h_{bc}\cdot \delta h^{ac}
-\frac{1}{2}h^{ab}D_c D_d\delta h_{ab}\cdot \delta h^{cd} \non
&& +\frac{1}{2}D^c\delta h^{ab}\cdot D_c\delta h_{ab}-2h^{ab}D^c\delta h_{ac}\cdot D^d\delta h_{bd}
+h^{ab}D^c\delta h_{ab}\cdot D^d\delta h_{cd}+D^c\delta h^{ab}\cdot D_b\delta h_{ac} \,,
\eean
and
\bean
-h^{ab}\delta(D^cD_c\delta h_{ab}) \eqn -\delta[D^cD_c(h^{ab}\delta h_{ab})]+\delta h^{ab}\cdot D^cD_c\delta h_{ab} \non
\eqn -[D_c\delta(D^c(h^{ab}\delta h_{ab}))+\delta C^c_{cd}D^d(h^{ab}\delta h_{ab})]+\delta h^{ab}\Box \delta h_{ab} \non
\eqn -h^{ab}D_c\delta h^{cd}\cdot D_d\delta h_{ab}-\Box(\delta h^{ab}\delta h_{ab})-h^{ab}\Box\delta^2 h_{ab} \non
&& -\frac{1}{2}h^{ab}h^{cd}D^e\delta h_{ab}D_e\delta h_{cd}+\delta h^{ab}\Box\delta h_{ab} \,.
\eean
So we obtain
\bean \label{genesecond}
\mathcal{G}_2 \eqn \int_C \frac{\chi\sgt}{2} \Big[2\delta h^{ac}D_c D^b\delta h_{ab}+2\delta h^{bd}D^a D_d\delta h_{ab}
+h^{ac}h^{bd}D_c D_d\delta^2 h_{ab}-3\delta h_{ab}\Box\delta h^{ab} \non
&& h^{ab}D_c D_d\delta h_{ab}\cdot \delta h^{cd}-\frac{3}{2}D^c\delta h^{ab}\cdot D_c\delta h_{ab}
-2h^{ab}D^c\delta h_{ac}\cdot D^d\delta h_{bd}+2h^{ab}D^c\delta h_{ab}\cdot D^d\delta h_{cd} \non
&& \left. +D^c\delta h^{ab}\cdot D_b\delta h_{ac}-h^{ab}\Box\delta^2 h_{ab}-\frac{1}{2}h^{ab}h^{cd}D^e\delta h_{ab}\cdot D_e\delta h_{cd}
+\rt_{ab}\delta^2h^{ab} \right] \non
&& +\chi\sgt (h^{ab}h^{cd}-h^{ac}h^{bd})u^{e}u^{f}(\grad_a\delta g_{eb}+\grad_b\delta g_{ae}-\grad_{e}\delta g_{ab})\cdot
(\grad_c\delta g_{fd}+\grad_d\delta g_{cf}-\grad_{f}\delta g_{cd}) \,.
\eean
While the third term of \eq{delta2S} gives
\bean \label{genethird}
\mathcal{G}_3 \eqn \int_C\frac{1}{12}\chi\left(\rt h^{cd}+2R_{ef}h^{ce}h^{df}-Rh^{cd}\right)h_{cd}\sgt\delta h^{ab}\cdot\delta h_{ab} \non
&& +\frac{1}{4}\chi\left(\rt h^{ab}+2R_{cd}h^{ac}h^{bd}-R h^{ab}\right)\sgt\delta^2 h_{ab} \non
&& +\frac{1}{8}\chi\left(\rt h^{ab}+2R_{ef}h^{ae}h^{bf}-R h^{ab}\right)\sgt h^{cd}\delta h_{ab}\cdot\delta h_{cd} \,.
\eean
Note that some relations satisfied in background spacetime could simplified the calculation of the fourth term, such as \cite{fang1}
\bean
(p+\rho)h^{ab} = \rt\hsp^{ab}-A^aA^b-D^bA^a+h^{ab}\grad_c A^c \,,
\eean
which yields $3(p+\rho)=\rt+2\grad_aA^a$. So the fourth term of \eq{delta2S} becomes
\bean
\mathcal{G}_4 \eqn \int_C -\frac{\sgt}{T}\frac{\delta p\delta\rho}{p+\rho} \non
\eqn \int_C -\frac{3\chi\sgt}{2(\rt+2\grad_cA^c)}\delta \rt\cdot[\delta(p h_{ab})-p\delta h_{ab}]h^{ab} \,.
\eean
The standard calculation yields \cite{waldbook}
\bean
\delta\rt \eqn h^{ab}\delta\rt_{ab}+\rt_{ab}\delta h^{ab} \non
\eqn D^aD^b\delta h_{ab}-h^{bc}\Box\delta h_{bc}-\rt\hsp^{ab}\delta h_{ab} \,.
\eean
And we have
\bean
h^{ab}[\delta(p h_{ab})-p\delta h_{ab}]
\eqn h^{ab}\delta(R^{cd}h_{ac}h_{bd}-\frac{1}{2}Rh_{ab})-(R_{cd}h^{ac}h^{bd}-\frac{1}{2}Rh^{ab})\delta h_{ab} \non
\eqn 2R^{ab}\delta h_{ab}-h^{ab}\delta R_{ab}-\frac{3}{2}\delta R-R_{cd}h^{ac}h^{bd}\delta h_{ab} \non
\eqn 2R^{ab}\delta h_{ab}-[\delta(R_{ab}h^{ab})-R_{ab}\delta h^{ab}]-\frac{3}{2}\delta R-R_{cd}h^{ac}h^{bd}\delta h_{ab} \non
\eqn R^{ab}\delta h_{ab}-\frac{1}{2}\delta(\rt-R)-\frac{5}{2}\delta R-R_{cd}h^{ac}h^{bd}\delta h_{ab} \non
\eqn R^{ab}\delta h_{ab}-\frac{1}{2}D^aD^b\delta h_{ab}+\frac{1}{2}h^{bc}\Box\delta h_{bc}+\frac{1}{2}\rt\hsp^{ab}\delta h_{ab}
-R_{cd}h^{ac}h^{bd}\delta h_{ab} \non
&& -2\grad^{a}\grad^{b}\delta g_{ab}+2g^{bc}\grad^{a}\grad_{a}\delta g_{bc}+2R^{ab}\delta g_{ab} \,,
\eean
where the variation of \eq{rtR} has been used. Then the fourth term can be written as
\bean \label{genefourth}
\mathcal{G}_4 \eqn -\frac{3\chi\sgt}{4(\rt+2\grad_hA^h)}(D^aD^b\delta h_{ab}-h^{ab}\Box\delta h_{ab}-\rt\hsp^{ab}\delta h_{ab})
\cdot \Big[2R^{cd}\delta h_{cd}-2R_{ef}h^{ce}h^{df}\delta h_{cd} \non
&& -D^cD^d\delta h_{cd}+h^{cd}\Box\delta h_{cd}+\rt\hsp^{cd}\delta h_{cd}-4\grad^{c}\grad^{d}\delta g_{cd}
+4g^{cd}\grad^{e}\grad_{e}\delta g_{cd}+4R^{cd}\delta g_{cd}\Big] \,.
\eean

Note that the state under perturbation is deviated only slightly from equilibrium state, the $\delta^2h_{ab}$ terms in the expression of $\delta^2S$ should vanish. In fact, denoting the sum of all terms containing $\delta^2h_{ab}$ by $\mathcal{G}_{\delta^2h_{ab}}$, we have
\bean
\mathcal{G}_{\delta^2 h_{ab}} \eqn \int_C\frac{\chi\sgt}{2}[h^{ac}h^{bd}D_c D_d\delta^2 h_{ab}
-h^{ab}\Box\delta^2 h_{ab}+\rt_{ab}\delta^2h^{ab}] \non
&& +\frac{\chi\sgt}{4}\left(\rt h^{ab}+2R_{cd}h^{ac}h^{bd}-R h^{ab}\right)\delta^2 h_{ab} \non
\eqn \int_C\frac{\sgt}{2}\left\{-D_b D_a\chi+h_{ab}\Box\chi+\chi\rt_{ab}
+\frac{\chi}{2}\left(-\rt h_{ab}-2R_{cd}h_a\hsp^ch_b\hsp^d+R h_{ab}\right) \right\}\delta^2h^{ab} \non
\eqn \int_C\frac{\sgt}{2}\left[-D_b(\chi A_a)+h_{ab}D^c(\chi A_c)+\chi R_{aeb}\hsp^lu^eu_l-\chi R_{cd}u^cu^d h_{ab}\right]\delta^2h_{ab} \,.
\eean
In the last step, we use the relation $D_a\chi=\chi A_a$ in static background spacetime. Combining the fact \cite{fang1}
\bean
\grad_au_b=-A_bu_a \,,
\eean
it could be proved that
\bean
\mathcal{G}_{\delta^2 h_{ab}}\equiv 0 \,.
\eean
Therefore, substituting \eqs{genefirst}, \meq{genesecond}, \meq{genethird} and \meq{genefourth} into \eq{delta2S}, we have
\bean \label{delta2Smid}
\delta^2S \eqn \int_C\frac{\chi\sgt}{2}\Big[h^{cd}\delta h_{cd}\cdot D^aD^b\delta h_{ab}
-h^{ab}h^{cd}\delta h_{cd}\cdot \Box\delta h_{ab}-h^{cd}\rt\hsp^{ab}\delta h_{ab}\delta h_{cd}+2\delta h^{ac}\cdot D_c D^b\delta h_{ab} \non
&& +2\delta h^{bd}\cdot D^a D_d\delta h_{ab}-3\delta h_{ab}\cdot \Box\delta h^{ab}
+h^{ab}D_c D_d\delta h_{ab}\cdot \delta h^{cd}-\frac{3}{2}D^c\delta h^{ab}\cdot D_c\delta h_{ab} \non
&& -2h^{ab}D^c\delta h_{ac}\cdot D^d\delta h_{bd}+2h^{ab}D^c\delta h_{ab}\cdot D^d\delta h_{cd}
+D^c\delta h^{ab}\cdot D_b\delta h_{ac}-\frac{1}{2}h^{ab}h^{cd}D^e\delta h_{ab}\cdot D_e\delta h_{cd} \Big] \non
&& +\chi\sgt (h^{ab}h^{cd}-h^{ac}h^{bd})u^{e}u^{f}(\grad_a\delta g_{eb}+\grad_b\delta g_{ae}-\grad_{e}\delta g_{ab})\cdot
(\grad_c\delta g_{fd}+\grad_d\delta g_{cf}-\grad_{f}\delta g_{cd}) \non
&& +\frac{\chi\sgt}{12}\left(\rt h^{cd}+2R_{ef}h^{ce}h^{df}-Rh^{cd}\right)h_{cd} \cdot\delta h^{ab}\delta h_{ab} \non
&& +\frac{\chi\sgt}{8}\left(\rt h^{ab}+2R_{ef}h^{ae}h^{bf}-R h^{ab}\right)h^{cd}\delta h_{ab}\cdot\delta h_{cd} \non
&& -\frac{3\chi\sgt}{4(\rt+2\grad_hA^h)}(D^aD^b\delta h_{ab}-h^{ab}\Box\delta h_{ab}-\rt\hsp^{ab}\delta h_{ab})
\cdot \Big[2R^{cd}\delta h_{cd}-2R_{ef}h^{ce}h^{df}\delta h_{cd} \non
&& -D^cD^d\delta h_{cd}+h^{cd}\Box\delta h_{cd}+\rt\hsp^{cd}\delta h_{cd}-4\grad^{c}\grad^{d}\delta g_{cd}
+4g^{cd}\grad^{e}\grad_{e}\delta g_{cd}+4R^{cd}\delta g_{cd}\Big] \,,
\eean

It should note that for static background spacetime, $g_{0i}=0$ (here the index $i=1,2,3$). However, the system may not remain static after the perturbation, which means that the perturbation fields including $\delta g_{00}$, $\delta g_{0i}$ and $\delta h_{ij}=\delta g_{ij}$. Now we decompose $\delta g_{ab}$ into $\delta g_{00}$, $\delta g_{0i}$ and $\delta h_{ij}$. The term $-4g^{ab}\grad^e\grad_e\delta g_{ab}$ in \eq{delta2Smid} can be calculated as
\bean \label{50}
&& -4\grad^c\grad^d\delta g_{cd} = -4g^{ca}\pp_a(g^{db}\grad_b\delta g_{cd})+4g^{ca}\Gamma^b\hsp_{ac}g^{de}\grad_e\delta g_{bd} \non
\eqn -4g^{ca}\pp_a[g^{db}(\pp_b\delta g_{cd}-\Gamma^e\hsp_{bc}\delta g_{ed}-\Gamma^e\hsp_{bd}\delta g_{ce})] +4g^{ca}g^{de}\Gamma^b\hsp_{ac}(\pp_e\delta g_{bd}-\Gamma^f\hsp_{eb}\delta g_{fd}-\Gamma^f\hsp_{ed}\delta g_{bf}) \,.
\eean
Then $\delta g_{ab}$ can be decomposed into $\delta g_{00}$, $\delta g_{0i}$ and $\delta h_{ij}$ (See Appendix A).
And the term $4g^{cd}\grad^{e}\grad_{e}\delta g_{cd}$ in \eq{delta2Smid} can be written as
\bean
4g^{cd}\grad^{e}\grad_{e}\delta g_{cd} = 4\grad^e\grad_e(g^{cd}\delta g_{cd})
= 4\grad^e\grad_e(g^{00}\delta g_{00}+h^{ij}\delta h_{ij}) \,.
\eean
It can be proved that $R^{0i}=0$ in background static spacetime, so
\bean
4R^{cd}\delta g_{cd}=4R^{00}\delta g_{00}+4R^{ij}\delta h_{ij} \,.
\eean
Finally, the expression of $\delta^2S$ can be written as
\bean \label{delta2Sresult}
\delta^2S \eqn \int_C\frac{\chi\sgt}{2}\Big[h^{cd}\delta h_{cd}\cdot D^aD^b\delta h_{ab}
-h^{ab}h^{cd}\delta h_{cd}\cdot \Box\delta h_{ab}-h^{cd}\rt\hsp^{ab}\delta h_{ab}\delta h_{cd}+2\delta h^{ac}\cdot D_c D^b\delta h_{ab} \non
&& +2\delta h^{bd}\cdot D^a D_d\delta h_{ab}-3\delta h_{ab}\cdot \Box\delta h^{ab}
+h^{ab}D_c D_d\delta h_{ab}\cdot \delta h^{cd}-\frac{3}{2}D^c\delta h^{ab}\cdot D_c\delta h_{ab} \non
&& -2h^{ab}D^c\delta h_{ac}\cdot D^d\delta h_{bd}+2h^{ab}D^c\delta h_{ab}\cdot D^d\delta h_{cd}
+D^c\delta h^{ab}\cdot D_b\delta h_{ac}-\frac{1}{2}h^{ab}h^{cd}D^e\delta h_{ab}\cdot D_e\delta h_{cd} \Big] \non
&& +\frac{\sgt}{\chi} (h^{ab}h^{cd}-h^{ac}h^{bd})(\pp_a\delta g_{0b}+\pp_b\delta g_{a0}-\pp_{0}\delta h_{ab}-2\Gamma^{i}\hsp_{ab}\delta g_{0i}) (\pp_c\delta g_{0d}+\pp_d\delta g_{c0}-\pp_{0}\delta h_{cd}-2\Gamma^{i}\hsp_{cd}\delta g_{0i}) \non
&& +\frac{\chi\sgt}{12}\left(\rt h^{cd}+2R_{ef}h^{ce}h^{df}-Rh^{cd}\right)h_{cd} \cdot\delta h^{ab}\delta h_{ab} \non
&& +\frac{\chi\sgt}{8}\left(\rt h^{ab}+2R_{ef}h^{ae}h^{bf}-R h^{ab}\right)h^{cd} \cdot\delta h_{ab}\delta h_{cd} \non
&& -\frac{3\chi\sgt}{4(\rt+2\grad_hA^h)}(D^aD^b\delta h_{ab}-h^{ab}\Box\delta h_{ab}-\rt\hsp^{ab}\delta h_{ab})
\cdot \Big[2R^{cd}\delta h_{cd}-2R_{ef}h^{ce}h^{df}\delta h_{cd}-D^cD^d\delta h_{cd} \non
&& +h^{cd}\Box\delta h_{cd}+\rt\hsp^{cd}\delta h_{cd}+4\grad^{c}\grad_{c}(g^{00}\delta g_{00})
+4\grad^{c}\grad_{c}(h^{ij}\delta h_{ij})+4R^{00}\delta g_{00}+4R^{ij}\delta h_{ij}-4g^{ab}\grad^e\grad_e\delta g_{ab}\Big] \,, \non
\eean
where the last term of \eq{delta2Sresult} can be written as
\bean
&& -4g^{ab}\grad^e\grad_e\delta g_{ab} \non
\eqn -4(g^{00})\hsp^2\pp_0^2\delta g_{00}+4h^{ij}\pp_j(g^{00}\Gamma^0\hsp_{0i}\delta g_{00})
-4(g^{00})\hsp^2\Gamma^i\hsp_{00}\Gamma^0\hsp_{0i}\delta g_{00}
-4g^{00}h^{jk}\Gamma^i\hsp_{jk}\Gamma^0\hsp_{0i}\delta g_{00} \non
&& -4g^{00}h^{ij}\pp_0\pp_j\delta g_{0i}-4h^{ij}\pp_j(g^{00}\pp_0\delta g_{0i})
+4g^{00}h^{ij}\Gamma^0\hsp_{j0}\pp_0\delta g_{0i} \non
&& +12(g^{00})\hsp^2\Gamma^i\hsp_{00}\pp_0\delta g_{0i}+8g^{00}h^{jk}\Gamma^i\hsp_{jk}\pp_0\delta g_{0i} \non
&& -4h^{ik}\pp_k(h^{jl}\pp_l\delta h_{ij})+4h^{lm}\pp_m(h^{jk}\Gamma^i\hsp_{kl}\delta h_{ij})
+4h^{ik}\pp_k(g^{00}\Gamma^j\hsp_{00}\delta h_{ij})+4h^{ik}\pp_k(h^{lm}\Gamma^j\hsp_{lm}\delta h_{ij}) \non
&& +4g^{00}h^{jk}\Gamma^i\hsp_{00}\pp_k\delta h_{ij}+4h^{lm}h^{jk}\Gamma^i\hsp_{lm}\pp_k\delta h_{ij}
-4g^{00}h^{jk}\Gamma^l\hsp_{00}\Gamma^i\hsp_{kl}\delta h_{ij}-4h^{mn}h^{jk}\Gamma^l\hsp_{mn}\Gamma^i\hsp_{kl}\delta h_{ij} \non
&& -4(g^{00})\hsp^2\Gamma^i\hsp_{00}\Gamma^j\hsp_{00}\delta h_{ij}-8g^{00}h^{kl}\Gamma^i\hsp_{00}\Gamma^j\hsp_{kl}\delta h_{ij}
-4h^{kl}h^{mn}\Gamma^i\hsp_{kl}\Gamma^j\hsp_{mn}\delta h_{ij}
\eean
\end{widetext}

It is shown that whether the system is stable depends on $\delta^2S<0$ under the perturbation $\delta g_{\mu\nu}$. This result gives the criterion for thermodynamical stability for perfect fluid star in static background without spherical symmetry. It is worth noting that this result also gives the criterion for non-radial perturbations cases.

\section{Summary and discussion}

Maximum entropy principle suggests a close relations between thermodynamics and gravity. In previous paper, we obtained the first variation of total entropy for perfect fluid in static spacetime and proved that the Einstein equations could be derived from the extrema of total entropy and the constraint equation with some boundary conditions \cite{fang1}. That is to say, the gravitational equations could be replaced by thermodynamical relations and constraint equation.

In this manuscript, we investigate the thermodynamical stability of an adiabatic, self-gravitating perfect fluid system deviated only slightly from equilibrium state. With thermodynamical relations, we obtain the expression of the second variation of total entropy and the criterion for thermodynamical stability. Specific to Einstein's gravity with spherical symmetry spacetime and radial perturbation, we give the explicit expression of our criterion and show that it is the same as the one in \cite{waldgeneral} which was obtained by dynamical method. For more general cases without spherical symmetry, we transform all variation of thermodynamical quantities to the variation of geometry quantities. Considering perfect fluid system in a static background spacetime, our criterion could be used directly to determine whether the system is stable under any specified perturbations. Our result establishes a connection between thermodynamic and gravity in higher order variation.

Using dynamical method, it is hard to handle the stability problems of general cases without spherical symmetry or under non-radial perturbations. However, in the framework of thermodynamical method, the stability only depends on the signature of $\delta^2S$. Furthermore, if the Lagrangian for diffeomorphism invariant theories is constructed by metric and its symmetrised derivatives, the criterion for thermodynamical stability \eq{delta2S} could also be used in this modified theories, such as $f(R)$ theories. In fact, we also proved that the thermodynamical stability is equivalent to dynamical stability in $f(R)$ theories \cite{equivafR}. And we found that using thermodynamical method to obtain the stability criterion is much more directly than dynamical method. Note that if the Lagrangian contains other scalar or vector parts, \eq{deltaS} need to be modified \cite{fang2}, which yields that the criterion for thermodynamical stability also need to be modified.

\acknowledgments
We thank Sijie Gao and Zhoujian Cao for many useful discussions and comments
on the manuscript. Jing was supported by the NSFC (No.~11475061). Fang was
supported by the NSFC (No.~11705053). He was supported by the NSFC (No.~11401199).

\begin{widetext}
\begin{appendix}
\section{Decomposition of $\delta g_{ab}$ in \eq{50}}

In this appendix, we show how to decompose $\delta g_{ab}$ in \eq{50} into $\delta g_{00}$, $\delta g_{0i}$ and $\delta h_{ij}$. In the following calculation the fact that $g^{0i}=0$ and $\Gamma^0\hsp_{00}=\Gamma^0\hsp_{ij}=\Gamma^i\hsp_{j0}=0$ in static background spacetime would be used. Then we calculate the term $-4\grad^c\grad^d\delta g_{cd}$, which can be written as
\bean \label{gcgdgcd}
&& -4\grad^c\grad^d\delta g_{cd}  \non
\eqn -4g^{ca}\pp_a[g^{db}(\pp_b\delta g_{cd}-\Gamma^e\hsp_{bc}\delta g_{ed}-\Gamma^e\hsp_{bd}\delta g_{ce})]
+4g^{ca}g^{de}\Gamma^b\hsp_{ac}(\pp_e\delta g_{bd}-\Gamma^f\hsp_{eb}\delta g_{fd}-\Gamma^f\hsp_{ed}\delta g_{bf}) \,,
\eean
The first term of \eq{gcgdgcd} can be calculated as
\bean \label{gcgd1}
&& -4g^{ca}\pp_a[g^{db}(\pp_b\delta g_{cd})] \non
\eqn -4g^{00}\pp_0(g^{00}\pp_0\delta g_{00})-4g^{00}\pp_0(g^{ij}\pp_j\delta g_{0i})
-4g^{ij}\pp_j(g^{00}\pp_0\delta g_{i0})-4g^{ik}\pp_k(g^{jl}\pp_l\delta g_{ij}) \non
\eqn -4(g^{00})\hsp^2\pp_0^2\delta g_{00}-4g^{00}h^{ij}\pp_0\pp_j\delta g_{0i}
-4h^{ij}\pp_j(g^{00}\pp_0\delta g_{i0})-4h^{ik}\pp_k(h^{jl}\pp_l\delta h_{ij}) \,.
\eean
Similarly, the left terms of \eq{gcgdgcd} can be written as
\bean \label{gcgd2}
&& -4g^{ca}\pp_a[g^{db}(-\Gamma^a\hsp_{bc}\delta g_{ed})] \non
\eqn 4h^{ij}\pp_j(g^{00}\Gamma^0\hsp_{0i}\delta g_{00})+4g^{00}h^{ij}\Gamma^0\hsp_{j0}\pp_0\delta g_{0i}
+4(g^{00})\hsp^2\Gamma^i\hsp_{00}\pp_0\delta g_{i0}+4h^{lm}\pp_m(h^{jk}\Gamma^i\hsp_{kl}\delta h_{ij}) \,,
\eean
\bean \label{gcgd3}
&& -4g^{ca}\pp_a[g^{db}(-\Gamma^e\hsp_{bd}\delta g_{ce})] \non
\eqn 4(g^{00})\hsp^2\Gamma^i\hsp_{00}\pp_0\delta g_{0i}+4g^{00}h^{jk}\Gamma^i\hsp_{jk}\pp_0\delta g_{0i}
+4h^{ik}\pp_k(g^{00}\Gamma^j\hsp_{00}\delta h_{ij})+4h^{ik}\pp_k(h^{lm}\Gamma^j\hsp_{lm}\delta h_{ij}) \,,
\eean
\bean \label{gcgd4}
&& 4g^{ca}g^{de}\Gamma^b\hsp_{ac}\pp_e\delta g_{bd} \non
\eqn 4(g^{00})\hsp^{2}\Gamma^i\hsp_{00}\pp_0\delta g_{i0}+4g^{00}h^{jk}\Gamma^i\hsp_{jk}\pp_0\delta g_{i0}
+4g^{00}h^{jk}\Gamma^i\hsp_{00}\pp_k\delta h_{ij}+4h^{lm}h^{jk}\Gamma^i\hsp_{lm}\pp_k\delta h_{ij} \,,
\eean
\bean \label{gcgd5}
&& 4g^{ca}g^{de}\Gamma^b\hsp_{ac}(-\Gamma^f\hsp_{eb}\delta g_{fd}) \non
\eqn -4(g^{00})\hsp^2\Gamma^i\hsp_{00}\Gamma^0\hsp_{0i}\delta g_{00}-4g^{00}h^{jk}\Gamma^i\hsp_{jk}\Gamma^0\hsp_{0i}\delta g_{00}
-4g^{00}h^{jk}\Gamma^l\hsp_{00}\Gamma^i\hsp_{kl}\delta h_{ij}-4h^{mn}h^{jk}\Gamma^l\hsp_{mn}\Gamma^i\hsp_{kl}\delta h_{ij} \,,
\eean
\bean \label{gcgd6}
&& 4g^{ca}g^{de}\Gamma^b\hsp_{ac}(-\Gamma^f\hsp_{ed}\delta g_{bf}) \non
\eqn -4(g^{00})\hsp^2\Gamma^i\hsp_{00}\Gamma^j\hsp_{00}\delta h_{ij}-8g^{00}h^{kl}\Gamma^i\hsp_{00}\Gamma^j\hsp_{kl}\delta h_{ij}
-4h^{kl}h^{mn}\Gamma^i\hsp_{kl}\Gamma^j\hsp_{mn}\delta h_{ij} \,.
\eean
Together with \eqs{gcgd1} $\sim$ \meq{gcgd6}, after some calculation, we have
\bean
&& -4\grad^c\grad^d\delta g_{ab} \non
\eqn -4(g^{00})\hsp^2\pp_0^2\delta g_{00}+4h^{ij}\pp_j(g^{00}\Gamma^0\hsp_{0i}\delta g_{00})
-4(g^{00})\hsp^2\Gamma^i\hsp_{00}\Gamma^0\hsp_{0i}\delta g_{00}
-4g^{00}h^{jk}\Gamma^i\hsp_{jk}\Gamma^0\hsp_{0i}\delta g_{00} \non
&& -4g^{00}h^{ij}\pp_0\pp_j\delta g_{0i}-4h^{ij}\pp_j(g^{00}\pp_0\delta g_{0i})
+4g^{00}h^{ij}\Gamma^0\hsp_{j0}\pp_0\delta g_{0i} \non
&& +12(g^{00})\hsp^2\Gamma^i\hsp_{00}\pp_0\delta g_{0i}+8g^{00}h^{jk}\Gamma^i\hsp_{jk}\pp_0\delta g_{0i} \non
&& -4h^{ik}\pp_k(h^{jl}\pp_l\delta h_{ij})+4h^{lm}\pp_m(h^{jk}\Gamma^i\hsp_{kl}\delta h_{ij})
+4h^{ik}\pp_k(g^{00}\Gamma^j\hsp_{00}\delta h_{ij})+4h^{ik}\pp_k(h^{lm}\Gamma^j\hsp_{lm}\delta h_{ij}) \non
&& +4g^{00}h^{jk}\Gamma^i\hsp_{00}\pp_k\delta h_{ij}+4h^{lm}h^{jk}\Gamma^i\hsp_{lm}\pp_k\delta h_{ij}
-4g^{00}h^{jk}\Gamma^l\hsp_{00}\Gamma^i\hsp_{kl}\delta h_{ij}-4h^{mn}h^{jk}\Gamma^l\hsp_{mn}\Gamma^i\hsp_{kl}\delta h_{ij} \non
&& -4(g^{00})\hsp^2\Gamma^i\hsp_{00}\Gamma^j\hsp_{00}\delta h_{ij}-8g^{00}h^{kl}\Gamma^i\hsp_{00}\Gamma^j\hsp_{kl}\delta h_{ij}
-4h^{kl}h^{mn}\Gamma^i\hsp_{kl}\Gamma^j\hsp_{mn}\delta h_{ij}
\eean

\end{appendix}
\end{widetext}

\end{document}